\documentclass[journal]{IEEEtran}
\usepackage{amsmath}
\usepackage{amssymb}
\usepackage{enumerate}
\usepackage{commath}
\usepackage{color}
\usepackage{graphicx}
\usepackage{multirow}
\usepackage{bigstrut}
\usepackage{tabularx}
\usepackage[flushleft]{threeparttable}
\usepackage{mathtools}
\usepackage{multicol}

\DeclarePairedDelimiter \floor{\lfloor}{\rfloor}

\ifCLASSINFOpdf
\else
\fi
\begin{document}

\bstctlcite{IEEEexample:BSTcontrol}

%
\title{Detecting the Presence of ENF \textcolor{black}{Signal} in Digital Videos: a Superpixel based Approach}
%
%
%
\author{Saffet~Vatansever,~
        Ahmet~Emir~Dirik,~
        Nasir Memon
\thanks{This material is based on research sponsored by DARPA and the Air Force Research Laboratory (AFRL) under agreement number FA8750-16-2-0173. The U.S. Government is authorized to reproduce and distribute reprints for Governmental purposes notwithstanding any copyright notation thereon. The views and conclusions contained herein are those of the authors and should not be interpreted as necessarily representing the official policies or endorsements, either expressed or implied, of DARPA and the Air Force Research Laboratory (AFRL) or the U.S. Government.}        
\thanks{S. Vatansever is with the Department
of Mechatronics Engineering, Bursa Technical University , Bursa 16333, Turkey,
and also with Uludag University, Bursa 16059, Turkey.}
\thanks{A. E. Dirik (Corresponding Author) is with the Department of Computer Engineering, Faculty of Engineering, Uludag University, Bursa 16059, Turkey (e-mail: edirik@uludag.edu.tr).}
\thanks{N. Memon is with the Department of Computer Science and Engineering, Tandon School of Engineering, New York University, Brooklyn, NY 11201 USA.}}

%
%

\markboth{IEEE SIGNAL PROCESSİNG LETTERS}%
{Shell \MakeLowercase{\textit{et al.}}: Bare Demo of IEEEtran.cls for IEEE Journals}
%



\maketitle

\begin{abstract}

ENF (Electrical Network Frequency) instantaneously fluctuates around its nominal value (50/60 Hz) due to a continuous disparity between generated power and consumed power. Consequently, \textcolor{black}{luminous} intensity of a mains-powered light source varies depending on ENF fluctuations in the grid network. \textcolor{black}{Variations in the luminance over time can be captured from video recordings and ENF can be estimated through content analysis of these recordings. In ENF based video forensics, it is critical to check whether a given video file is appropriate for this type of analysis.} That is, if ENF signal is not present in a given video, it would be useless to apply ENF based forensic analysis. In this work, an ENF signal presence detection method is introduced for videos. \textcolor{black}{The proposed method is based on multiple ENF signal estimations from steady superpixels, i.e. pixels that are most likely uniform in color, brightness, and texture, and intra-class similarity of the estimated signals}. Subsequently, consistency among these estimates is then used to determine the presence or absence of an ENF signal in a given video. The proposed technique can operate on video clips as short as  2 minutes and is independent  of the camera sensor type, i.e. CCD or CMOS.

%
\end{abstract}


%
\IEEEpeerreviewmaketitle

\section{Introduction}
%
%
%
%

%
%

%
%

\IEEEPARstart{E}{NF} (Electrical Network Frequency) is a time varying signal fluctuating continuously around its nominal value (50/60 Hz) due to the instantaneous imbalance between power consumption and power generation \cite{Bollen2006}. For each time instance, ENF fluctuation is almost the same across the entire interconnected power grid network \cite{Grigoras2005}. Accordingly, electric frequency measured at any location connected to a particular mains power can be used as a reference ENF signal for the whole area covered by that power network for the relevant time period \cite{Garg2011}. This property of electric frequency, as well  as the ability to extract it from multimedia files, has led to the exploitation of ENF in digital media forensics in recent years. \textcolor{black}{It can be used for a variety of \textcolor{black}{forensic and anti-forensic} applications including audio/video authentication \cite{Hua2016, Savari2016}, \textcolor{black}{\cite{Chuang2013}}, time stamp verification \cite{Fechner2014, Bykhovsky2013, MinWu-SeeingENFPower-signature-basedtimestampfordigitalmultimediaviaopticalsensingandsignalprocessing}, and power grid identification \cite{MinWu-ENF-BasedRegion-of-RecordingIdentificationforMediaSignals}.} 

An ENF signal is embedded in audio recordings made in settings where the electromagnetic field or acoustic mains hum exists and it can be estimated from these recordings with time-domain or frequency-domain approaches \cite{Grigoras2005}, \cite{Fechner2014}, \cite{grigoras2007}, \cite{Grigoras2009}, \cite{Chai2013}. Recently, it was found that ENF can also be estimated from video captured under  illumination of a light source powered by mains grid \cite{Garg2011}. The intensity of illumination from any light source connected to the mains power varies depending on ENF variations in the grid network. Although the human eye cannot perceive, ENF can be estimated by analysis of subtle illumination variations in steady content along subsequent video frames.

Two different ENF estimation methods  for ENF based forensic analysis of digital video \textcolor{black}{have been} proposed in the literature: a method tailored to videos recorded by CCD sensor \textcolor{black}{\cite{Garg2011}} and a technique for videos captured by CMOS sensor \cite{Garg2011, MinWu-SeeingENFPower-signature-basedtimestampfordigitalmultimediaviaopticalsensingandsignalprocessing, Hajj-Ahmad2016a}. While the former  is based on averaging all the steady pixels in each frame along the video, the latter processes steady pixels based on a rolling shutter sampling mechanism \cite{Hajj-Ahmad2016a}, \cite{MinWu-ENFSignalInducedbyPowerGridANewModalityforVideoSynchronization}, \cite{MinWu-ExploringtheuseofENFformultimediasynchronization}, \cite{Su2014}.


In ENF based video forensics, it is \textcolor{black}{important} to test whether a video contains any traces of ENF  before moving on to further analysis. \textcolor{black}{For instance}, if a video does not contain \textcolor{black}{any} ENF signal it would be useless to \textcolor{black}{search in existing ENF databases for video \textcolor{black}{time-stamp or region-of-recording verification}}. \textcolor{black}{More importantly,} a substantial amount of computational load and time can be saved if a quick test can establish the absence of an ENF signal. To the best of our knowledge, none of the work in the literature presents an approach that can automatically detect the presence of an ENF signal in a video \textcolor{black}{regardless of the imaging sensor type, e.g. CCD or CMOS.} 

In this letter, a superpixel based ENF signal presence detection technique is proposed. The proposed method performs multiple ``so-called ENF'' signal estimations from different steady object regions having very close reflectance properties, i.e. superpixels \cite{Achanta2010}. \textcolor{black}{Our motivation to use superpixels is that each pixel in a superpixel region is almost uniform in brightness, color and texture, and hence has uniform reflectance characteristics. Working on such a region provides the possibility of estimating ENF from videos taken by not only CCD camera but also by CMOS camera, which uses rolling shutter mechanism.} \textcolor{black}{In the proposed algorithm, a ``so-called ENF signal" is estimated from each steady superpixel separately.} \textcolor{black}{The reason we use the term ``so-called ENF" \textcolor{black} is {because} \textcolor{black}{the estimated signal} is initially unknown to be actually an ENF signal}. \textcolor{black}{Depending on the similarity of the estimated signals from each steady superpixel, it can be decided whether any ENF signal is  present in the test video or not.} 
It should be noted  that the proposed method does not require any \textcolor{black}{verification}  \textcolor{black}{against} a reference ENF database. 

\section{ENF Power Model}
\label{Sec: ENF Power Model}
Instantaneous power grid voltage can be modeled as follows:
\begin{align}
\begin{split}
V(t) &= \textcolor{black}{\sqrt{2} V_0 \cos(\phi(t))} \\ 
       &= \sqrt{2} V_0 \cos(2\pi f_{n}t+\theta(t)+\textcolor{black}{\alpha})\\
       &=\sqrt{2} V_0 \cos(2\pi f_{n}t+2\pi\int_0^t {f_e(\tau)\mathrm{d}\tau }+\textcolor{black}{\alpha})\\
\end{split}
\label{Eq_PowerGridVoltage}
\end{align}
where $f_n$ is nominal frequency (50/60Hz), \textcolor{black}{$V_0$} is effective mains voltage and \textcolor{black}{$\alpha$} is initial phase offset \cite{Bollen2006}.
$f_e(t)$ represents instantaneous fluctuations from nominal frequency and $\theta(t)$ denotes instantaneous phase which varies depending on supply-demand power imbalance. From the above equations, the instantaneous mains power frequency at time $t$ can be expressed as
\begin{align}
\begin{split}
f(t) &=\frac{1}{2\pi}\frac{d\textcolor{black}{\phi(t)}}{dt}=f_{n}+f_e(t)\\ 
\end{split}
\end{align}
As $f_n$ is constant, electric network frequency alters depending on $f_e(t)$ variations. By benefiting from \textcolor{black}{the model in} \cite{Bollen2006}, $f_e(t)$ can be written as
\begin{align}
f_e(t) =\frac{f_n}{2 H}(P_s(t)-P_d(t)) 
\end{align}
where $P_s(t)$ denotes supplied power, $P_d(t)$ is total demanded power with losses and $H$ represents an inertia constant. Accordingly, for each time instance, $f_e(t)$ and  $f(t)$ \textcolor{black}{change} depending on the instantaneous difference between generated and consumed power.

\section{\textcolor{black}{ENF Estimation from Video}}
\label{Sec: A Robust ENF Estimation Method from Videos}
\subsection{\textcolor{black}{Light Source Flicker and ENF}}
\label{sec:ENF on Light Source Illumination}

Intensity of illumination from any light source connected to the mains power varies depending on ENF variations in grid network. As light source flickers at both the positive and negative cycles of AC current, the illumination frequency \textcolor{black}{becomes} double the mains power frequency. Accordingly, the illumination signal can be treated as the absolute form of the cosine function in $(\ref{Eq_PowerGridVoltage})$. \textcolor{black}{For example in Europe,} 
nominal ENF in any region is 50 Hz,  \textcolor{black}{thus} frequency of illumination varies around 100 Hz. 
\textcolor{black}{According to the Nyquist Sampling Theorem, a sampling rate of at least 200 \textcolor{black}{Hz} is needed in order to extract illumination frequency accurately from \textcolor{black}{sampled} data. \textcolor{black}{Although most consumer cameras are unable to provide such high frame sampling rates, it is still possible to estimate illumination frequency from its alias frequency.}
Let  $f_s$ be the camcorder sampling frequency and $f_l$ be the frequency of light source illumination. Then  $f_a$  the aliased frequency of illumination  is obtained as follows \cite{Hajj-Ahmad2015}:
\begin{align}
f_a=\abs{f_l-k \cdot f_s} < \frac{f_s}{2}, \quad \exists k \in \mathbb{\textcolor{black}{N}}
\end{align}
Accordingly, when a light source illumination signal in 100 Hz is sampled with 29.97 fps camera, the base alias frequency of ENF is obtained as 10.09 Hz. 
}




\subsection{Superpixel based ENF Estimation}
\label{SubSec: superpixel based ENF estimation}

\begin{figure}[!t]
\centering
\includegraphics[height=40mm]{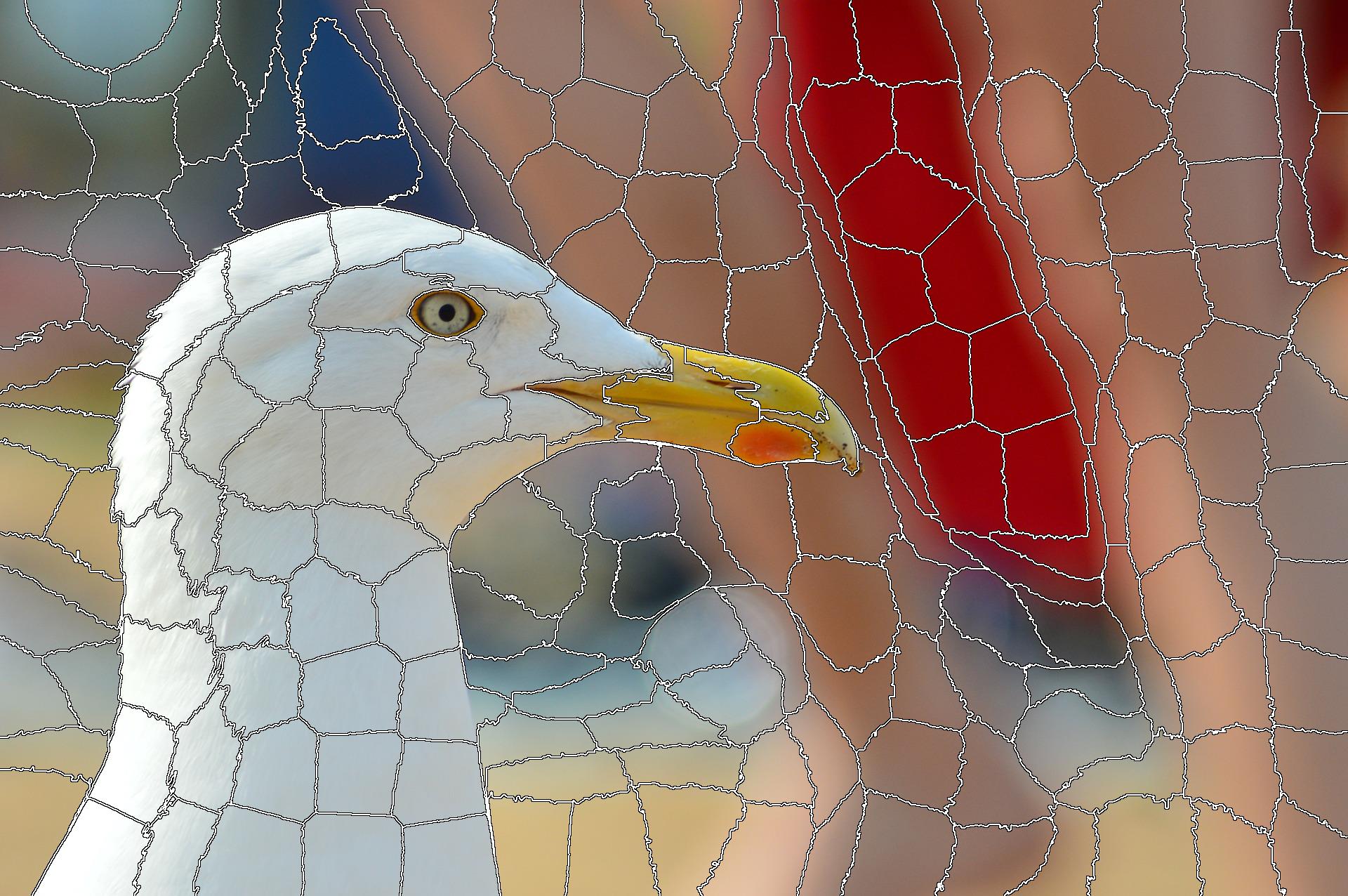}
\caption{A sample image with superpixels.}
\label{Fig: A sample figure with SLIC segmentation}
\end{figure}

In this section, a \textcolor{black}{superpixel based} ENF estimation method for video is proposed. Unlike available methods in the literature \cite{Garg2011}, \cite{MinWu-SeeingENFPower-signature-basedtimestampfordigitalmultimediaviaopticalsensingandsignalprocessing}, \textcolor{black}{we propose to estimate ENF from only steady superpixels, rather than all steady pixels along video frames.}
The SLIC (Simple Linear Iterative Clustering) segmentation algorithm \cite{Achanta2010} is used to compute superpixel regions in the experiments. \textcolor{black}{A sample image segmented with the SLIC algorithm and its superpixels is given in Fig. \ref{Fig: A sample figure with SLIC segmentation}}. The underlying idea of the proposed ENF estimation is that each pixel in a superpixel region/set is assumed to have uniform reflectance characteristics.

The amount of illumination at any pair of pixel coordinates $x,y$, received by a camera at any moment $n$ can be \textcolor{black}{written} as \cite{gonzalez2011digital}:
\begin{align}\label{eqI_1}
I(x,y,n) =i_{s}(x,y,n) \cdot r(x,y)
\end{align}
where $r(x,y)$ denotes the amount of \textcolor{black}{reflected illumination and $i_{s}(x,y,n)$ is instantaneous light source illumination. For a point light source, $i_{s}(x,y,n)$  can be expressed in terms of mains electricity voltage approximately as:}
\begin{align}\label{eqI_1b}
\textcolor{black}{i_s(x,y,n)\approx \frac{\beta}{d_s(x,y)^2} \cdot \abs{V(n)}}
\end{align}
where $\beta$ is a transform factor \textcolor{black}{for} voltage to luminance conversion \textcolor{black}{and $d_s(x,y)$ is the distance between the spatial position $(x,y)$ and the light source}. The reason  $V(n)$ is in absolute form is that light source \textcolor{black}{produces illumination in both positive and negative cycles of electrical power grid voltage}. \textcolor{black}{In a superpixel region, the reflectance factor $r(x,y)$ can be assumed to be constant}. Similarly, the distance of any pixel in a superpixel $S$ to the light source $d_s(x,y)$ can be considered constant as well. \textcolor{black}{Thus, for $k$th steady superpixel $S_k$, $(\ref{eqI_1})$ can be rewritten in the following form}:
\begin{align}\label{eqI_2}
\textcolor{black}{
I_k(x,y,n) \approx \beta \cdot \frac{r_k}{d_k^2} \cdot \abs{V(n)} \text{ , } (x,y)\in{S_k}}
\end{align}
\textcolor{black}{where $r_k$ denotes a constant reflectance factor for pixels belonging to $k$th steady superpixel region ($S_k$), $k \in \{1,...,L\}$. $L$ is the number of steady superpixels},  \textcolor{black}{$d_k$ is the approximate distance of superpixel $S_k$ to the light source.} As it can be seen from $(\ref{eqI_2})$, $I_k(x,y,n)$ is directly proportional to $V(n)$, which means that the frequency of power grid voltage $V(n)$ can be directly estimated from $I_k(x,y,n)$.
\textcolor{black}{Illumination variations at any superpixel region $S$ can be estimated by averaging the steady pixels in $S$, resulting in $L$ illumination vectors.}
\textcolor{black}{From each of the illumination vectors, ENF variations} \textcolor{black}{can be estimated using any of the frequency or time domain approaches discussed in \cite{MinWu-SeeingENFPower-signature-basedtimestampfordigitalmultimediaviaopticalsensingandsignalprocessing}, \cite{grigoras2007}. In this paper, ENF is estimated from local intensity variations using STFT (Short Time Fourier Transform) and quadratic interpolation. In order to compute STFT we have used 20 sec. time windows with 19 sec. overlapping resulting 1-second temporal ENF resolution \cite{Cooper2008}}. \textcolor{black}{It is important to note the content of the superpixel region should be unchanged all along the consecutive video frames, in order to estimate ENF successfully. This  will be addressed in the next section.}
%
%
%
%
%
\section{Detection of ENF \textcolor{black}{Signal} Presence}
\label{SubSec: ENF Presence Verification in Digital Videos}
In this section, a superpixel based ENF presence detector for digital video files is presented based on multiple \textcolor{black}{ENF signal} estimations from steady superpixel regions. The main steps of the proposed technique are illustrated in Table \ref{ENF detector}.
\textcolor{black}{According to the table, one frame, e.g. the middle frame $\bold{F}_r$, in a selected video shot $C$ is segmented into regions having similar pixel characteristics, i.e. superpixels.} 
Then, within each superpixel region, the points that are steady throughout all frames, i.e. non-moving pixels are located. \textcolor{black}{Superpixels having a low number of steady pixels ($m_l < \tau$) are not used in ENF estimation. \textcolor{black}{In this study, the value of $\tau$ has been determined empirically as 30$\times$30 pixels.} For each \emph{steady} superpixel $S_k$ and each video frame $\bold{F}_n$, the average intensity $Y_k (n)$ is computed from steady pixels of region $S_k$.
From each intensity variation vector $\bold{Y}_k$, a ``so-called" ENF vector $\bold{E}_k$ is estimated along all the subsequent video frames of the given shot.} As stated in introduction, the reason  we use the prefix ``so-called'  is that it is initially unknown to be actually ENF or not.

 \textcolor{black}{Next, the similarity of estimated ENF vectors is analyzed to decide whether ENF signal is present or not in the test video. For this purpose, a representative ENF vector $\bold{E}_r$ is computed by means of element-wise mean or median operation of all the estimated ENF vectors. Next, Pearson correlation coefficients between each estimated $\bold{E}_k$ and the representative vector $\bold{E}_r$ are calculated as follows:
\begin{align}
\begin{split}
\rho(k) &= \textrm{corr}(\bold{E}_k, \, \bold{E}_r)=\frac{<\bold{E}_k-\bar{\bold{E}}_k, \, \bold{E}_r-\bar{\bold{E}}_r>}{\parallel\bold{E}_k-\bar{\bold{E}}_k\parallel \,\parallel\bold{E}_r-\bar{\bold{E}}_r\parallel}\\
\end{split}
\label{Corr}
\end{align}
where $\parallel \cdot \parallel$ denotes $L_2$ (Euclidean) norm and $< \cdot >$ is the dot product. The sample mean is denoted with \textcolor{black}{overline}. Afterwards, a decision metric is computed based on the following operations: $f_1=\textrm{max}({\boldsymbol{\rho}})$, \, $f_2=\textrm{mean}(\boldsymbol{\rho})$, \, $f_3=\textrm{median}(\boldsymbol{\rho})$, \, $f_4=\textrm{corr}(\bold{E}_i, \, \bold{E}_j)$,
where $\bold{E}_i$ and $\bold{E}_{j}$ are the vectors yielding the greatest $\rho$ values (top two closest vectors to $\bold{E_r}$). If the decision metric is greater than a predefined decision threshold value, the video is labeled as having an ENF signal.}






\begin{table}[!t]
\renewcommand{\arraystretch}{1.3}
\caption{Algorithm: Detection of ENF Signal Presence}
\label{table_example}
\centering
\begin{tabular}{p{0.25cm} p{7.35cm}}
\hline\hline
\centering
\bfseries Step & \bfseries Description\\
\hline
1 & \textcolor{black}{Pick any video shot $C$. Let $\bold{F}_n$ be the $n$th frame in $C$, where $n\in \{1,...,N\}$.}\\

2 & \textcolor{black}{Let the middle frame $\bold{F}_r$ be the representative frame in $C$, where $r=\floor{N/2}$.}\\

3 & \textcolor{black}{Compute superpixel regions for the representative frame. Let $\Omega_l$ be the $l$th superpixel in $\bold{F}_r$, $l\in \{1,...,\textcolor{black}{P}\}$. $\textcolor{black}{P}$ is the total number of the superpixels.}\\

4 & \textcolor{black}{Let $\Phi$ be the set of all steady pixels in which the video content do not change  with time in $\bold{F}_r$.}\\

5 & \textcolor{black}{Compute the number of steady pixels $m_l$ in each $\Omega_l$ using the steady pixel set $\Phi$.}\\

6 & \textcolor{black}{Compute steady superpixel set $S$ from $\{\Omega_l\}$: $S=\{\Omega_l \mid m_l>\tau \}$, where $\tau$ is a pre-defined threshold for the minimum number of non-changing pixels in a superpixel region.}\\

7 & \textcolor{black}{For each steady superpixel $S_k$ and each frame $\bold{F}_n$, compute the average intensity $Y_k(n)$, only from steady pixels of region $S_k$. $k\in \{1,...,L\}$, and $L$ is the total number of \emph{steady} superpixels.}\\

8 & \textcolor{black}{Estimate ENF variation signal $\bold{E}_k$ from local intensity variations $\bold{Y}_k$ for each superpixel $S_k$.}\\

9 & \textcolor{black}{Place all $E_k(i)$ into a matrix $\bold{M}$ such that $M(k,i)=E_k(i)$, where $i \in \{1,...,t\}$ and $t$ is the ENF vector length. Let $\bold{E}_r$ be the representative ENF vector computed by means of element-wise median or mean operation of all $\bold{E}_k$ vectors, where $n$th sample of $\bold{E}_r$ is computed as:}\\
 & \textcolor{black}{$E_r(i|\text{mean}) = \underset{k}{\mathrm{mean}} \{M(k,i)\}$}\\
 & \textcolor{black}{$E_r(i|\text{median}) = \underset{k}{\mathrm{median}} \{M(k,i)\}$}\\
10 & \textcolor{black}{Compute similarity of each $\bold{E}_k$ with $\bold{E}_r$ by Pearson correlation, $\rho(k)$.}\\

11 & \textcolor{black}{Compute mean, median, maximum, and similar statistics of $\boldsymbol{\rho}$ vector as decision metrics.}\\

12 & \textcolor{black}{If the computed metric is greater than a predefined decision threshold, the presence of ENF signal in the video is confirmed.}\\

\hline\hline
\end{tabular}

\label{ENF detector}

\end{table}

\section{Experiments and Results}



\subsection{Experimental Setup}
\label{SubSec:ExperimentalSetup}
%
%
In this section, the performance of the proposed method is evaluated by conducting experiments on \textcolor{black}{partially-moving-content videos} captured in various indoor and outdoor settings in Turkey, where nominal ENF frequency is 50 Hz. \textcolor{black}{ENF signal presence} \textcolor{black}{was} searched \textcolor{black}{in} a total of \textcolor{black}{160 videos}, one half of which \textcolor{black}{were} recorded by PowerShot SX230HS (CMOS sensor) and the other half \textcolor{black}{were} recorded by Canon PowerShot SX210IS (CCD sensor). For CMOS, the Canon PowerShot SX230HS model camcorder \textcolor{black}{was} intentionally picked as it \textcolor{black}{has been} used in most ENF related works \cite{MinWu-SeeingENFPower-signature-basedtimestampfordigitalmultimediaviaopticalsensingandsignalprocessing}, \cite{Hajj-Ahmad2016a}, \cite{MinWu-ENFSignalInducedbyPowerGridANewModalityforVideoSynchronization} and \cite{Su2014}. The CCD equivalent of the same camera brand and model \textcolor{black}{was} picked so as to do a fair \textcolor{black}{comparison} of the algorithm according to the sensor type. \textcolor{black}{Out of 80 videos for each sensor-camera type, one-quarter was captured at night under illumination of various mains-powered light sources such as LED, fluorescent tube, CFL, tungsten halogen, sodium-vapor lamp, street light. A second quarter were recorded under illumination of mains-powered light sources but in daylight settings such as in a room with an opened window or next to a lamp on the balcony in the sunset afternoon. The third quarter were taken under illumination of non-mains-powered light sources in daylight settings and the last quarter were captured at night under illumination of non-mains-powered light sources such as moonlight, vehicle headlight, candle, projector torch, smart-phone torch and laptop screen}. Hence, each video is initially known  to contain an ENF signal or not. All the videos \textcolor{black}{were} created in 640$\times$480 resolution with a sampling frequency of 29.97 fps. \textcolor{black}{The camera was fixed during recording and it was ensured that each video has some steady content in addition to moving content.} When sampled with 29.97 fps, the peak alias of 100 Hz \textcolor{black}{can} be computed as 10.09 as discussed in section \ref{sec:ENF on Light Source Illumination}. Therefore, 10.09 Hz frequency band \textcolor{black}{was} utilized for ENF signal estimation from the relevant video files. Although the videos \textcolor{black}{were} in a variety of lengths between 2 minutes and 15 minutes, a 2-minute clip of each video was used for ENF detection experiment. It is also notable that the representative frame for each video \textcolor{black}{was} segmented into about 48 superpixel regions, which corresponds to about 6400 ($80\times80$) pixels per superpixel for a frame of 640$\times$480 pixels resolution.

%
%
\begin{figure}[t]
\centering
\includegraphics[width=80mm]{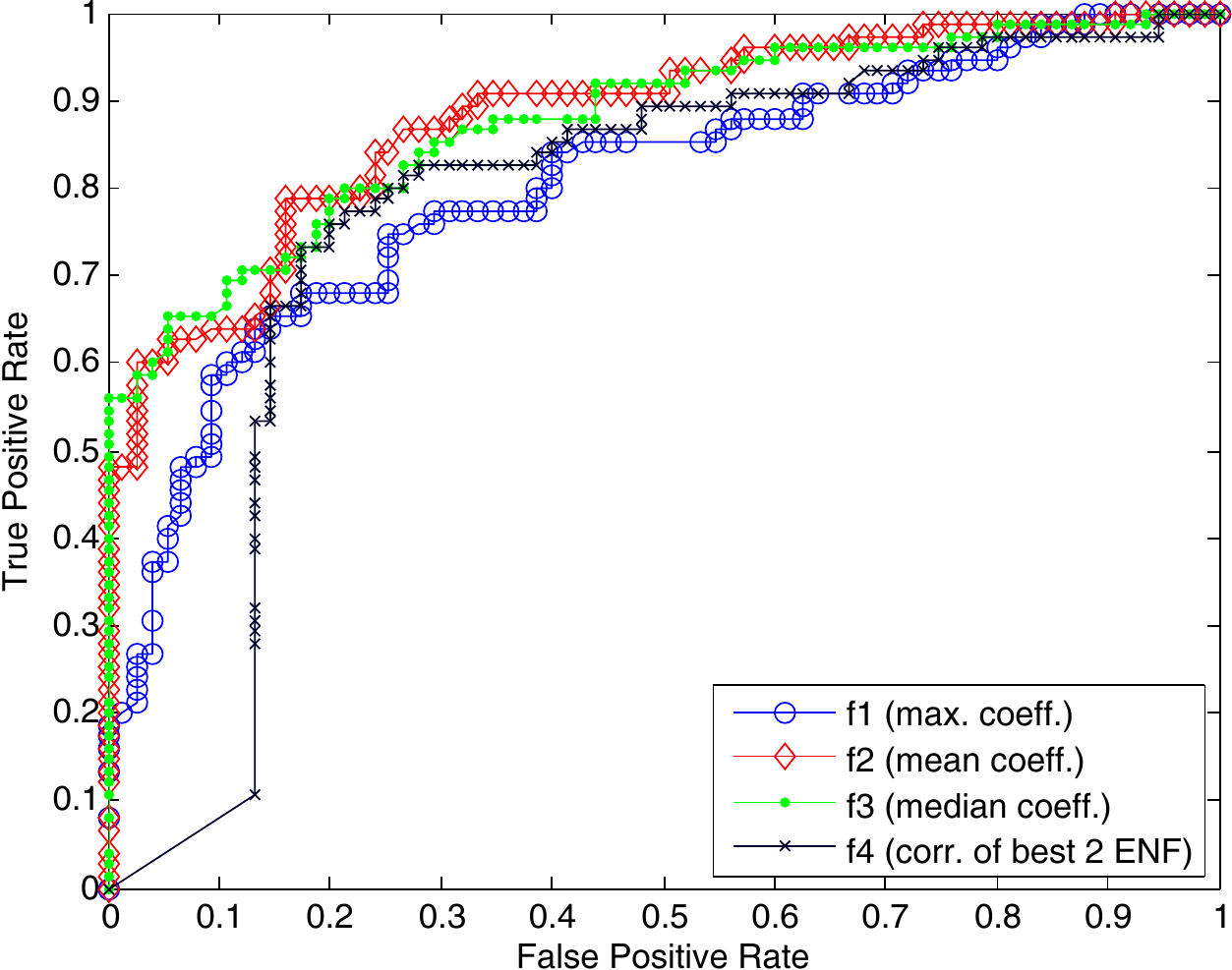}
\caption{\textcolor{black}{ROC curves of ENF presence detection for videos recorded by both CCD and CMOS sensors, \textcolor{black}{160} videos. $\bold{E}_r$ computed with $\bold{mean}$ operation.}}
\label{AnySensor_ROC_for_MeanRepresentative}
\end{figure}
\begin{figure}[t]
\centering
\includegraphics[width=80mm]{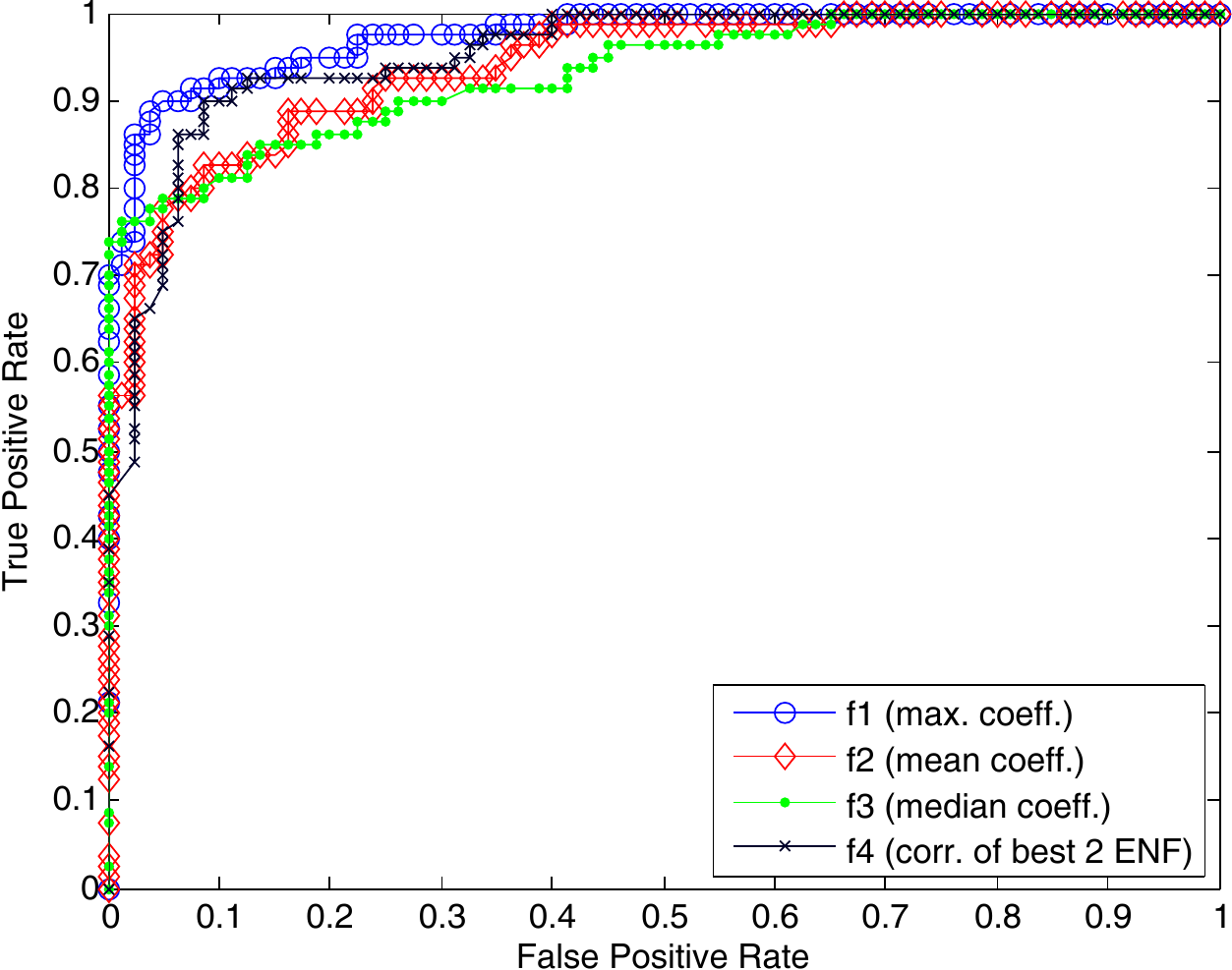}
\caption{\textcolor{black}{ROC curves of ENF presence detection for videos recorded by both CCD and CMOS sensors, \textcolor{black}{160} videos. $\bold{E}_r$ computed with the $\bold{median}$ operation.}}
\label{AnySensor_ROC_for_MedianRepresentative}
\end{figure}



\subsection{Experimental Results}

%

\begin{table}[t]
\caption{ENF Detection Performance (AUC) based on Mean based Representative ENF}
\centering
  \begin{tabular}{c c c c c c}
  	\hline\hline
	Sensor Type & \# Videos &  $f1$ & $f2$ & $f3$ & $f4$ \bigstrut\\
	\hline
	CCD & \textcolor{black}{80} & \textcolor{black}{0.836} & $\bold{\color{red}0.896}$ & \textcolor{black}{0.895} & \textcolor{black}{0.813} \bigstrut\\
    
	CMOS & \textcolor{black}{80} & \textcolor{black}{0.760} & $\bold{\color{red}0.883}$ & \textcolor{black}{0.866} & \textcolor{black}{0.700} \bigstrut\\
    
	Any (Mixed) & \textcolor{black}{160} & \textcolor{black}{0.798} & $\bold{\color{red}0.886}$ & \textcolor{black}{0.881} & \textcolor{black}{0.761} \bigstrut\\
    \hline\hline
  \end{tabular}
\label{Table: ENF Detection Rate-Mean based Representative ENF}
\end{table}

\begin{table}[t]
\caption{ENF Detection Performance (AUC) based on Median based Representative ENF}
\centering
  \begin{tabular}{c c c c c c}
  	\hline\hline
	Sensor Type & \# Videos &  $f1$ & $f2$ & $f3$ & $f4$ \bigstrut\\
	\hline
	CCD & \textcolor{black}{80} & $\bold{\color{red}0.985}$ & \textcolor{black}{0.947} & \textcolor{black}{0.931} & \textcolor{black}{0.960} \bigstrut\\
    
	CMOS & \textcolor{black}{80} & $\bold{\color{red}0.959}$ & \textcolor{black}{0.942} & \textcolor{black}{0.941} & \textcolor{black}{0.944} \bigstrut\\
    
	Any (Mixed) & \textcolor{black}{160} & $\bold{\color{red}0.973}$ & \textcolor{black}{0.939} & \textcolor{black}{0.931} & \textcolor{black}{0.952} \bigstrut\\
    \hline\hline
  \end{tabular}
\label{Table: ENF Detection Rate-Median based Representative ENF}
\end{table}
In this section, the accuracy of the proposed ENF detection algorithm is tested on the video dataset described in Section \ref{SubSec:ExperimentalSetup} by computation of a Receiver Operating Characteristics (ROC) curve and area under the curve (AUC). For this purpose, the following binary hypotheses are defined:\\
$H_0:$ The video does not contain ENF signal\\
$H_1:$ The video contains ENF signal\\
Under these hypotheses, $f_1$, $f_2$, $f_3$ and $f_4$ decision metrics, introduced in Section \ref{SubSec: ENF Presence Verification in Digital Videos}, \textcolor{black}{were} computed for each video. Each decision metric was computed with the use of both mean-based representative ENF and median-based representative ENF, respectively and was assigned to the corresponding hypothesis, $H_0$ or $H_1$. Fig. \ref{AnySensor_ROC_for_MeanRepresentative} provides ROC curves obtained for decision metrics which are formed via mean based representative ENF computation. Whereas Fig. \ref{AnySensor_ROC_for_MedianRepresentative} illustrates ROC curves obtained for decision metrics that are calculated based on the utilization of median based representative ENF. From the ROC curves in Fig. \ref{AnySensor_ROC_for_MeanRepresentative} and Fig. \ref{AnySensor_ROC_for_MedianRepresentative}, a significant enhancement in the detection performance can explicitly be observed for all metrics when the decision metrics are formed with the use of median-based representative ENF. Table \ref{Table: ENF Detection Rate-Mean based Representative ENF} and \ref{Table: ENF Detection Rate-Median based Representative ENF} provides the area under the ROC curves (AUC) in Fig. \ref{AnySensor_ROC_for_MeanRepresentative} and Fig. \ref{AnySensor_ROC_for_MedianRepresentative}, respectively as well as AUC values for each sensor type, separately. The computed AUC values in Table \ref{Table: ENF Detection Rate-Median based Representative ENF} are considerably higher not only for mixture of sensor types but also for each sensor type independently. According to the Fig. \ref{AnySensor_ROC_for_MedianRepresentative}, the detection metric $f_1$ outperforms other metrics when testing a mixture video dataset whose source sensor type is unknown.

\section{Discussion and Conclusion}

In this paper, a superpixel based ENF detection algorithm for video is presented. The proposed method is able to work on short video clips of about 2 minutes-length and can be used to detect and differentiate the videos that are appropriate for ENF based forensic analysis from ENF free videos on a disk under investigation or in social media. By doing this, ENF free videos are not exposed unnecessarily to the execution of entire ENF based analysis; hence a substantial amount of time and computational load can be saved. The algorithm is able to operate independently of the source camera sensor type, CCD or CMOS and achieves a very high ENF signal presence detection accuracy for videos captured by both sensor types. 
\ifCLASSOPTIONcaptionsoff
  \newpage
\fi



%

%
%

\bibliographystyle{IEEEtran}
\onecolumn
\begin{multicols}{2}
\bibliography{ENF}
\end{multicols}

%

%
%
%




\end{document}